# Structure and Dynamics of DNA-dendrimer complexation: Role of counterions, water and base pair sequence

## Prabal K. Maiti[1,a] and Biman Bagchi[2,b]

[1]Center for Condensed Matter Theory, Indian Institute of Science, Bangalore-12, India,
[2]Solid State and Structural Chemistry Unit, Indian Institute of Science, Bangalore -12, India.

## Abstract

**We study sequence dependent complexation between oligonucleotides (single strand DNA) and various generation ethylene diamine (EDA) cored poly amido amide (PAMAM) dendrimers through atomistic molecular dynamics simulations accompanied by free energy calculations and inherent structure determination. Simulations reveal formation of a stable complex and provide a detailed molecular level understanding of the structure and dynamics of such a complexation. The reaction free energy surface in the initial stage is found to be funnel-like with a *significant* barrier arising in the late stage due to the occurrence of misfolded states of DNA. Complexation shows surprisingly strong sensitivity to the ssDNA sequence which is found to arise from a competition between enthalpic versus entropic rigidity of ssDNA.**

---

[a] maiti@physics.iisc.ernet.in
[b] bbagchi@sscu.iisc.ernet.in

# I.    Introduction

Understanding complexation and association reaction between two large macromolecules is a subject of great interest in physics, chemistry and biology because such reactions are ubiquitous in nature. Examples include macroion-polyelectrolyte complex, protein-DNA interaction, protein association, and DNA-dendrimer complexation. The interaction of polyelectrolyte with oppositely charged macroion (such as colloids) is important in several technological applications such as paper making, producing pigment coatings to name a few. The interaction of DNA with protein is another striking example of macroion-polyelectrolyte complex. In cell nucleus DNA is wrapped around positively charged protein know as histone and forms the nucleosome structure. These nucleosomes form higher order structure like beads on a string and produce chromatin structure [1]. However, no microscopic study could yet be carried out to explore microscopic picture of DNA coiling around protein because the histone exists in an octameric form and the problem is many orders of magnitude more difficult.  Polyelectrolyte-macroion complex are also attractive from a fundamental point of view as they give rise to other interesting problems like like-charge attraction due to structural correlation, overcharging of macroion with oppositely charged polyelectrolyte [2] and polyelectrolyte multilayering [3] for which there exists only few theoretical and numerical works. DNA-dendrimer complexation provides one of the simplest examples of such reactions, especially if we consider complexation between a single strand DNA and a charged dendrimer, such a PAMAM dendrimer. Here the driving force is largely electrostatic. At physiological pH, PAMAM dendrimer is positively charged. Therefore, they can effectively bind negatively charged DNA. Interest in this problem has increased in recent years because of the bio-medical applications that such a complex could provide. The complex can be used as gene delivery material inside mammalian cells [4-6]. There are indications that lower toxicity and higher transfection efficiency can be obtained by using complexes between DNA and PAMAM dendrimers [7].

 Because of the relative simplicity of DNA-dendrimer complexation reaction, one hopes to achieve a quantitative understanding of this process at a microscopic level.  Here the questions concern the pathway and mechanism of the complexation, the reaction coordinates that determine the free energy surface of the reaction, the relative role energy and entropy, how the environment and other various factors such as pH, salt concentration affects the DNA-dendrimer binding and the resultant conformation. The complexity of any such reaction (other well-known example is protein-DNA complexation [8, 9] or protein association [10, 11]) arises from the involvement of many degrees of freedom, expected to be highly cooperative, involving conformational change not only of the DNA, but also of water and the counter-ions. The present study provides the first microscopic details of this condensation. The present system of DNA-dendrimer binding not only addresses fundamental issues involved in the above mentioned complex biological reactions, it also helps to elucidate some of the finer details of macroion-polyelectrolyte and/or colloids polyelectrolyte complex at a microscopic details as PAMAM dendrimer can be viewed as a compact colloidal particle [12]. We are not yet aware of any study, which investigated free energy surface and reaction pathway for such complex many-body reactions.

A number of theoretical and computer simulation studies have been reported on structural properties of polyelectrolyte-dendrimer complexation under various conditions [13-16]. However, all these studies have focused on simple bead models of dendrimers and polyelectrolyte chains to obtain qualitative features. These studies have provided increased insight into some structural properties of dendrimer-polyelectrolyte systems. Welch and Muthukumar[14] first reported the complexation between a model dendrimer (chemically closer to Poly-propyl- Imine (PPI) dendrimer) with charged terminal groups and charged liner chains under varying pH conditions using Monte Carlo (MC) simulations. They also predicted theoretically adsorption/desorption criteria depending on the salt concentration, size of the dendrimer, charge density of dendrimer and polyelectrolyte chain and length of the linear polymer and found good agreement with the simulation results. The electrostatic interactions were treated within Debye-Huckel approximation. More recently using Brownian Dynamics simulation Lyulin et. al. [15] have reported the structural aspects of the complexes formed by charged dendrimer and oppositely charged linear polymer. They have reported the overcharging phenomena in the complex in accordance with the theoretical predictions [17]. In a series of paper Netz and co-workers [18-20] have studied the complex formation between a macroion and oppositely charged polyelectrolyte using linearized Possion-Boltzmann theory and reported various configuration of the complex depending on the ionic strength and valency of the macroion. In these calculations effects of counterions were taken into account implicitly and solvent structure was ignored. In all these studies solvent was treated as a continuum dielectric medium of constant dielectric constant. Role of water in the dendrimer dynamics as well as it's binding to ions and DNA has been established earlier through atomistic MD simulations [21, 22]. So the important role of solvent structure is missing from all the current theoretical and numerical studies of macroion and polyelectrolyte complex. Also none of the above studies take into account the DNA like properties such as base specificity, base stacking, hydrogen bonding and sequence dependence in the calculations. Missing also is the effect of explicit ions, which are known to play important role in DNA and dendrimer dynamics. A reliable resolution to such issues requires application of a fully atomistic description of dendrimers as well as DNA including explicit water and counterions. In this paper we investigate the complexation PAMAM dendrimers of generation 2-4 (G2-G4) at various protonation levels (structure given in **Figure 1 in the supplementary materials**) and 38 base pairs ssDNA (sequence given in **Figure 2 in the supplementary materials**) in explicit water and counterions using fully atomistic simulation. To study the sequence dependence we have also performed simulations with different sequences of ssDNA. We also report some results on the complexation between a double-stranded DNA (dsDNA) and dendrimer.

## II. Simulation Details

All MD simulations reported in this paper used the AMBER7 software package [23] with the all-atom AMBER95 force field (FF) [24] for DNA and Dreiding force field for dendrimer. The initial structure of G2-G4 PAMAM dendrimers at various protonation levels was taken from our previous study [22, 25]. At intermediate or neutral pH (pH ~ 7) all the primary amines (16 for G2, 32 for G3 and 64 for G4) are protonated. At low pH (pH ~ 4) all the primary amines and tertiary amines (14 for G2, 30 for G3 and 62 for G4) are protonated. For ssDNA structure we took one of the strand of the fully equilibrated 38 base pairs double strand B-DNA from our earlier work [26, 27]. Using the LEAP module in AMBER, dendrimers of various generations at various protonation levels were put in the major groove of ssDNA. The resulting structure was immersed in a water box using the TIP3P model for water [28]. The box dimensions were chosen in order to ensure a 10Å solvation shell around the DNA-dendrimer structure. In addition, some waters were replaced by $Na^+$ counter ions to neutralize the negative charge on the phosphate groups of the backbone of the DNA structures. Then appropriate numbers of Cl- ions were added to neutralize the positive charges on the dendrimer amine sites. This procedure resulted in solvated structures, containing approximately 74,000 atoms.

The solvated structures were then subjected conjugate gradient minimization. The electrostatics interactions were calculated with the Particle Mesh Ewald (PME) method [29, 30] using a cubic B-spline interpolation of order 4 and a $10^{-4}$ tolerance set for the direct space sum cutoff. A real space cut off of 9Å was used both for the electrostatics and van-der Waals interactions with a non-bond list update frequency of 10.

During the minimization the DNA and dendrimer structures were fixed in their starting conformations using harmonic constraints with a force constant of 500 kcal/mol/$Å^2$. This allowed the water molecules to reorganize to eliminate bad contacts with the DNA and dendrimer structures. The minimized structures were then subjected to 40 ps of MD, using a 2 fs time step for integration. During the MD, the system was gradually heated from 0 to 300 K using weak 20 kcal/mol/ $Å^2$ harmonic constraints on the solute to its starting structure. This allows slow relaxation of the built DNA-dendrimer structure. In addition SHAKE constraints [31] using a geometrical tolerance of 5 x$10^{-4}$ Å were imposed on all covalent bonds involving hydrogen atoms. Subsequently, MD was performed under constant pressure - constant temperature conditions (NPT), with temperature regulation achieved using the Berendsen weak coupling method [32] (0.5 ps time constant for heat bath coupling and 0.2 ps pressure relaxation time).

Finally, for analysis of structures and properties, we carried out 20 ns of NVT MD using a heat bath coupling time constant of 1 ps. All the simulations were carried our on 32 CPU SGI-Altix server and IBM Regatta server and took about 6 months of computing time.

# III Numerical Results and Theoretical Analysis

## III.1 Structural aspects of the complex:

<u>Size of the complex</u>

Calculating the mean squared radius of gyration $\left\langle R_g^2 \right\rangle$ monitored the conformational change of dendrimer, DNA in a complex as well as the dendrimer-DNA complex as whole. In **figure 1** we show the evolution of the mean squared radius of gyration of ssDNA in complex with G4 PAMAM dendrimer at neutral pH. At neutral pH G2, G3 and G4 PAMAM have 16, 32 and 64 protonated amines respectively while ssDNA has 37 negative charges. For G2 and G3 the charge ratio is not enough to have a complete wrapping of the ssDNA onto dendrimer as is evident from the larger value of the radius of gyration of the complex as well the ssDNA (see Fig. 3 in the supporting information). Only at higher generation like G4 when dendrimer charge is enough to neutralize the ssDNA charge do we see collapse of ssDNA on the surface of dendrimer and we achieve a very compact complex. For smaller dendrimer like G2 and G3 we see coiling of ssDNA in the vicinity of dendrimer only. Away from dendrimer ssDNA is in stretched conformations. For higher generation dendrimer with larger size with more positive charge ssDNA not only wraps around dendrimer, we see also significant penetration inside dendrimer. This is evident from the snapshot shown in **Fig. 4 in the supporting information** as well as from the average radial density distribution functions shown in **Figure 2.**

The average radial monomer density ρ(r) can be defined by counting the number of atoms $N(R)$ whose centers of mass are located within the spherical shell of radius r and thickness Δr. Hence, the integration over r yields the total number of atoms as:

$$N(R) = 4\pi \int_0^R r^2 \rho(r) dr$$

**Figure 2 show** the monomer radial distribution functions for dendrimer, DNA and water when DNA binds to the dendrimer for three different generations of dendrimer (G2-G4) with different charges and wrapping pattern. Several features deserve attentions: (1) Due to the swelling of dendrimer significant water penetrates inside dendrimer and the minima in the dendrimer density near the core correspond to the increase water density inside. Location of this minima does not change in going from G2 to G4 indicting the fact that first hydration layer inside the dendrimer remain at distance 10 Å from the center of mass of the dendrimer. More and more water penetrate inside with increase in the size of the dendrimer (going from G2 to G4). (2) As the ssDNA binds to the charged protonated amine sites of the dendrimer ssDNA monomer distribution shows maximum near the maxima in the location of the dendrimer terminal groups. Locations of these maxima shifts to the higher values with the increase in dendrimer size and roughly correspond to the radius of gyration of dendrimer. (3) DNA binding results in the expulsion of water from the region of terminal groups, which is manifested by appearance of minima in the water density profile at this location. At neutral pH we see high degree of DNA penetration inside the dendrimer and as the DNA goes inside dendrimer significant number of water is expelled from dendrimer and water density distribution functions shows a dip where DNA distribution functions shows it maxima. (4) Compared to G2 and G3, for G4 we find almost whole ssDNA monomer distribution inside the dendrimer density distribution function indicating complete penetration of the DNA inside dendrimer. This degree of penetration increases with the increase in the size of dendrimer or increase in the dna-dendrimer charge ratio. This implies that when the complex is formed too much of DNA penetration may complicate its release and thus making its use as gene therapy material difficult. However this chain penetration can be tuned by changing pH as has been discussed in section III.4. This is an important aspects coming out of this study. Our fully atomistic MD simulations bring out beautifully such detailed molecular level view of the complexation for the first time.

## Location of counterions and issues of Overcharging or Undercharging

In all the previous studies of model-based simulation on DNA-polymer complexes crucial role of counterions has not been addressed as the effect of the ions were included through Debye-Hückel approximation. In our simulation we have explicit ions and water included which allow us to systematically extract the ion-atmosphere around the dendrimer as well surrounding the DNA-dendrimer complex. A significant amount of counter ions penetrate in the interior of the dendrimer, which in turn affect the dynamics of the complexation.

Correlation theory predicts [17] a complete wrapping of an oppositely charged worm-like chain onto a sphere when the total charge on the chain equal to the charge of the sphere. However, in our simulation we see very different scenario. Complete wrapping occurs when the total charge on the chain is almost *1.5* times the charge of dendrimer. When the total charge on the chain equal to the charge of the dendrimer (as in the case of G3) we see partial adsorption and appearance of tail regions in the complex. Strong adsorption is observed for larger generation dendrimer when ssDNA charge is significantly smaller than the dendrimer charge. The reason of this discrepancy between our simulation results and correlation theory is the neglect of the effect of explicit solvent and counterions as well as the hydrogen bond.

In **figure 3** we show the total charge of the complex as a function of the distance from the center of mass of the complex for various generations of dendrimers. At small distance from the center of mass of the complex total charge is zero as no part of the ssDNA or charged terminal groups or counterions penetrates up to that distance. The extent of zero charge regime decreases with increase in the dendrimer generation (increase in the DNA-dendrimer charge ratio) as more and more ssDNA fragment penetrates the complex. Also with increasing dendrimer generation back-folding of the charged primary amine groups increases which means a significant number of them can be found in the interior of the complex even within 10 Å from the center of mass. After the zero charge regime the total charge passes through a positive maxima due to the back folding of the charged primary amine groups. Finally it passes through positive maxima and the location of the maxima corresponds to the radius of gyration of the dendrimer where the maximum in the distribution function of the terminal group occurs. The positive charge of the complex indicates overcharging phenomena. The overcharging of DNA-wrapped dendrimer complex has been found experimentally for G2 dendrimer [33].

### III.2 Dynamics of the Complex Formation

MD simulations reveal that it takes 2 ns for the ssDNA to start wrapping around the DNA and in the next several nanoseconds (more than 12 ns) DNA overcomes *several energetic and entropic bottlenecks* to find its optimal wrapping patterns on the dendrimer surface. In (**Fig. 4 a**) we show several snapshots of the systems in a few ns interval to show the various stages in the DNA wrapping process. *Note the significant conformational change even after 13 ns*. The mean squared radius of gyration $\left\langle R_g^2 \right\rangle$ provides a measure of the conformational change of DNA (and thus serves as an order parameter) as it folds and wraps around the DNA. We have also monitored the size of the dendrimer during the simulation. These results are shown in **Fig. 1,** which shows that the dendrimer *expands noticeably in the early stage* in order to optimize electrostatic interactions with the DNA whose size decreases as it coils around the dendrimer. In (**Fig. 4b**) we show the time dependence of the number of contacts ($N_C$) between the DNA and the dendrimer. This figure shows interesting *oscillatory, almost intermittent binding dynamics*. $N_C$ also shows an interesting rise and sharp fall at early stage (**inset in Fig. 4b**) reflecting an initial cooperative disengagement between the DNA and the dendrimer in search for an optimum approach. By the end of 20 ns, DNA completely wraps the dendrimer and assumes the size of the dendrimer.

### III.3  Free energy surface

We have computed the free energy for the process of the wrapping along two reaction coordinates: the distance ($R_{DNA\text{-}DEN}$) between the center of mass of the dendrimer and DNA, and a local bending defined as inverse of $R_{LB}$ where $R_{LB} = \sqrt{\sum_{i=1}^{N_P} r_i^2}$ , where $r_i$ is the distance of the *i-th* phosphate site on the ssDNA from the center of the dendrimer and $N_P$ is the number of phosphate in the DNA backbone- this quantity is minimal for a chain wrapped around the surface of dendrimer, but is unbiased with respect to the conformation of the DNA on the dendrimer. (**Fig. 5 a**) shows the two-dimensional (with $R_{DNA\text{-}DEN}$ and $R_{LB}$ as the reaction coordinates) reaction free energy surface. Two deep minima are visible on the reaction pathway. The first, metastable free energy minimum is found at larger DNA-dendrimer distance ($\approx$20Å) and larger bending ($\approx$240 Å$^2$). The final stable minimum is found near $R_{DNA-DEN} \approx$ 15 Å and bending $\approx$ 180Å$^2$. In (**Fig. 5 b**) we show the projected (on $R_{LB}$) one-dimensional free energy curves versus local bend, $R_{LB}$. Bistable nature of the free energy surface is clear from this figure. The existence of the bound state as well as the free energy minima agrees well with the recent experimental results on similar system [34]. Inset in Fig**. 5 b** shows the free energy versus the center of mass separation $R_{DNA-DEN}$. In this coordinate, the free *energy landscape looks funnel-like, with a rugged landscape, somewhat akin to protein folding* [35, 36].

We find that the barrier in the folding process arises from the wrong loop formation associated with the base pairing in the ssDNA as well as stacking of the bases in these loop regions. Wrong paring between G-T, A-A and G-A base pairs and base stacking in these regions give rise to an *enthalpic bottleneck in the wrapping process*. To overcome this barrier DNA would require melting and unwrapping of loop regions. In (**Fig. 6a**), we show two such wrong base pair formation during the initial complexation. For these two base pairs, the phosphates adjacent to the bases come very close in the two loop regions during folding of the DNA. Such "wrong" base pair formation has been noticed earlier in the folding of ssDNA and has been analyzed earlier by using the rugged landscape picture popular in protein folding [37, 38]. Just as in the case of protein folding [39], the condensation process also exhibits multistage dynamics. It is also interesting to note that the barrier towards folding occurs towards the end of the process. In (**Fig. 6b**) we show the time evolution of the distance between the two base pairs. The sudden jump in the distance around 1.5 ns shows the breaking of the hydrogen bond forming a wrong base pair. Such detailed dynamical picture provides insight into the binding process.

### III.4 pH Dependence

At high pH the dendrimer is uncharged and the formation of any DNA-dendrimer complex is not seen. Instead, the pair moves further apart during the simulation (see figure 5 in the supporting information). At neutral pH, when the dendrimer is positively charged due to the protonation of all the primary amines, the strong electrostatics interaction helps the DNA strand collapse onto the dendrimer. This electrostatic attraction is resisted by the elastic energy of the ssDNA due to bending. And when the electrostatic energy overcomes the elastic energy of bending we see the collapse of the DNA onto dendrimer. Our simulation results support the model developed earlier by Manning et. al.[40, 41] while studying the interaction of polycation with dendrimers as well as binding of DNA to the histone octamer. As the DNA wraps onto dendrimer, the DNA gains in electrostatics energy but the dendrimer looses energy slightly. This is shown in (**Fig. 7a**). Compared to the neutral pH case we see less penetration of DNA at low pH. This is also evident from the snapshot shown in Figure 6 in the supporting information. This implies that at neutral pH when the complex is formed too much of DNA penetration may complicate its release and thus making its use as gene therapy material difficult. So low pH condition may be better suited for the purpose of the DNA delivery inside cell. This is an important aspects coming out of this study.

However, if electrostatics interaction is the only major driving force in the DNA wrapping process, then lowering the solution pH further (which increase the DNA-dendrimer charge ratio), should accelerate the wrapping, as various recent experiments have proposed [42]. But to the contrary we find that increasing the DNA-dendrimer charge ratio or lowering the solution pH *does not* necessarily guarantee the wrapping of DNA on the dendrimer. This is in agreement with the available theoretical and experimental results [18, 43] where various groups have reported the condition for DNA-histone/DNA-protein binding. At high salt concentration the positive charge on the dendrimer is neutralized by the accumulation of negative $Cl^-$ ions on the dendrimer surface as well in the interior of the dendrimer, which in turn reduce the propensity of the DNA dendrimer binding.

### III.5 Role of water and counterions

Water and counter ions play important role in the competition between the entropic loss and the enthalpic gain as the DNA collapse onto dendrimer. A spine of hydration seems to follow DNA bending closely, clearly showing the role of water in the binding process. In (**Fig. 7b**), we show the time evolution of the number of water molecules that are within 3Å within the DNA backbone. For comparison, we also plot the time dependence of the bending in **Fig. 7b (see inset).** One sees a clear correlation between the two – the water molecules move away from the DNA backbone as the bending increases. Note the intermittent dynamics in (**Figs. 7b**) – such intermittent dynamics have been observed earlier in the hydrogen bond rearrangement dynamics in water [44].

## IV. Sequence Dependence of Complexation

The complexation exhibits strong sequence dependence and the binding constant follows G>C>A>T> sequence (see Table 1). To explore this further, we calculated the entropy and the Helmholtz free energy difference of the ssDNA and dendrimer in complex state as well as for the free ssDNA and dendrimer using vibrational density of states analysis [45]. We find that polyG gains both in entropy and enthalpies in the bound state (relative to its free rod-like state), making the dendrimer-polyG complex thermodynamically the most stable state. For polyC although it looses conformational entropy marginally in the bound state, the enthalpic gain overcomes this loss and makes the dendrimer-polyC complex thermodynamically a stable state. However, for polyA and polyT the entropic and enthalpic balance is not achieved to give a free energy minimum. Thus, the entropy of the complexed DNA plays an important role in the stability of the complex. The surprising gain in entropy on binding for poly-G results from the unstacking of the bases upon binding to the dendrimer. Such increase in entropy of ssDNA has been observed earlier upon binding to protein [46, 47]. Also the entropic and enthalpic rigidity of the polyA and polyT sequence is well known from various single molecule fluorescence studies [48].

## V.  Single strand vs double strand

So far all our simulation was focused on the complexation of oligonucleotide and dendrimer. Question arises how the molecular picture of condensation differs when we have a dsDNA. To have a molecular understanding of the complexation of dsDNA and dendrimer we have also done series of simulation with 38 base pair long dsDNA with various generation dendrimer at neutral pH when the primary amines of the dendrimer are protonated. We see the complexation between the DNA and dendrimer. **Figure 7** in the supporting information shows a snapshot of the system showing the complexation between G3 PAMAM and 38 bp long dsDNA. We see the presence of dendrimer induces strong bending to the dsDNA. The binding mode is analogous to the first binding proposed by Chen et. al.[49].

## VI. Conclusion

Three aspects of the present work deserve special attention. First, we have shown that the complexed state is a free energy minimum and is strongly sequence dependent. This is significant because such a DNA-dendrimer complex must be stable enough to permit its use as a gene delivery material as well in other pharmaceutical applications (like antisense therapeutics). Second, the strong sequence dependence (justified on the basis of energy-entropy calculations) may allow novel use of dendrimer as a sequence analyzer. Third, the rugged nature of the free energy along the condensation pathway due to the occurrence of wrong base pairing is a novel finding, which can have important consequences.

We believe that the present study is the first theoretical demonstration of complexation between a DNA molecule and a dendrimer from a fully atomistic description. The following microscopic picture emerges from the present study. On initial approach, the dendrimer expands in order to increase the DNA-dendrimer surface contact. On subsequent close approach, the positive charge on the surface of dendrimer forces the collapse, which is resisted by the bending elasticity of the DNA. Once the latter is overcome, the coiling starts which also brings distant base pairs close together. At this state, pairing (by hydrogen bonds) between these distant pairs can occur. Such base pairing can lead to the formation of a metastable state, as shown by (**Fig. 5b**). However, this metastable state is un-favored, on two counts. First, it prevents further coiling. Second, it costs too much elastic energy. *The global minimum forms at lower bending but shorter distance of separation* (see **Figs. 5a-2b**). The free energy of the complexation should thus consist of sum of at least three terms: the bending energy, the electrostatic energy and the base pairing energy. The existing theoretical treatments have considered the first two (and also the excluded volume interactions), but not the base pairing contribution. It is the latter that gives rise to the ruggedness of the free energy landscape and brings the present problem close to protein folding where also wrong contact formation is partly responsible for the ruggedness or frustration in the free energy surface along the folding pathway.

**Acknowledgements:**

We thank Profs K. L. Sebastian and M. Glaser for useful discussions and Supercomputer Education and Research Center (SERC), IISc for generous computer time where all the computation has been carried out. This work was supported in parts by grants from DST and DBT, India.

## Figure Captions:

**Figure 1:** Evolution of the size of the DNA, dendrimer as well as of the complex showing the conformational change during the wrapping process. The simulations have been done for G4 PAMAM at neutral pH.

**Figure 2:** Density distribution for dendrimer, DNA and water in the complex. The distribution has been computed with respect to the center of mass of the dendrimer.

**Figure 3:** Total charge of the complex as a function of distance from the center of mass of the complex. The data has been averaged over 2000 configuration at each ps interval for 4 ns.

**Figure 4**: (a) Structure of ssDNA-dendrimer complex during various stages of the wrapping process at the interval of few ns. Dendrimer has been shown in the surface representation in white while the DNA is shown in the tube representation. (b) Variation of the number of contact points between DNA and dendrimer (any contact within 3 Å) from 1.5 ns to 9 ns showing the intermittency during the wrapping process; **(Inset)** Initial rise and subsequent lowering of contact point during first 500 ps of the dynamics. During the initial period as the dendrimer size grows due to swelling arising out of electrostatic interactions, the DNA is pushed further lowering the number of contact points.

**Figure 5:** (a) Free energy contour map versus the DNA-dendrimer center of mass distance ($R_{DNA-DEN}$) and local bending ($R_{LB}$). The free energy landscape is determined by calculating the normalized probability [50], $P(X) = Z^{-1} \exp(-\beta W(X))$, where $X$ is any set of reaction coordinates. The relative free energy or the potential of mean force (PMF) has been calculated along $R_{DNA-DEN}$ and $R_{LB}$ as $F(X_2) - F(X_1) = -RT \ln\left(\dfrac{P(X_2)}{P(X_1)}\right)$

(b) 1-d free energy map versus the local bending coordinate showing the barrier in the wrapping process; **(Inset)** 1-d free energy map versus the center of mass separation of DNA and dendrimer ($R_{DNA-DEN}$).

**Figure 6:** (a) Snapshot showing the formation of two-loop region during the dynamics which gives rise to the barrier in the wrapping process. Shown are (green dotted lines) also the various hydrogen bonding involved in the mis-pairing of bases as well as base stacking. The loop formation occurs in between 1.5 –5 ns, which give rise to the barrier in the wrapping process. (b) Inter Phosphate distance in the loop region showing how close the bases come together to form the loops.

**Figure 7:** (a) Variation of the electrostatics energy of DNA and dendrimer during the binding process. As the DNA wraps around dendrimer it gains electrostatic energy, which offsets the elastic bending energy. On the other hand many of the water molecules, which were solvating the dendrimer as well as counter ions surrounding the dendrimer are pushed out as the DNA wraps and in the process effectively dendrimer become more positively charged. So the electrostatic energy of the dendrimer becomes more positive.
 (b) No water molecules in a spine of hydration (within 3 Å of the DNA backbone) as a function of time. Notice that as the local bending increase due to the wrapping more and more water molecules are expelled from the spine of hydration. **(Inset)** Local bend variation as a function of time as the DNA bends and wraps around dendrimer; To calculate the local bend we have calculated the contour length ($s$) of the DNA during the dynamics and the bending angle is defined as $\theta = \dfrac{s}{R_{LB}}$.

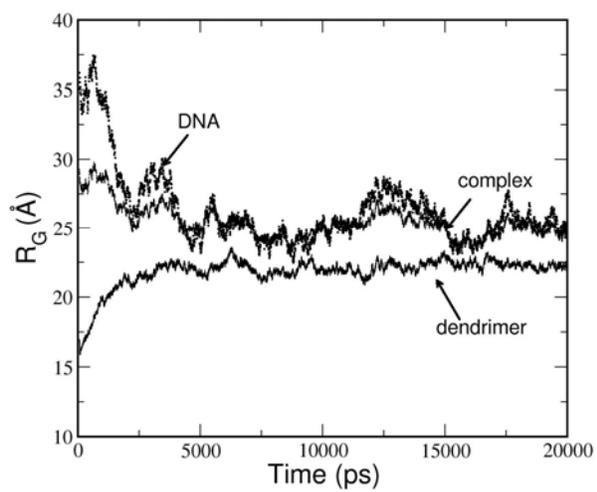

**Figure 1:**

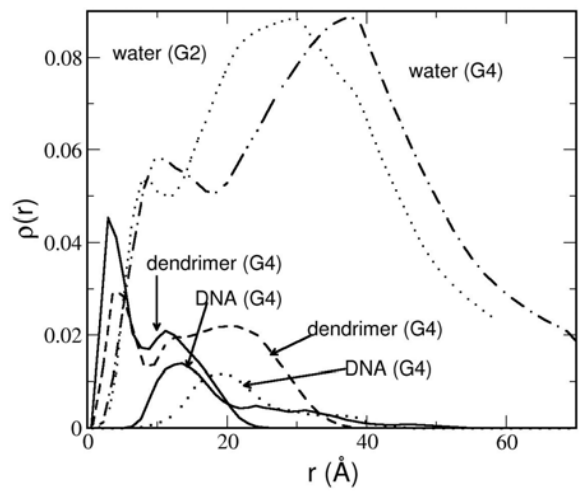

**Figure 2:**

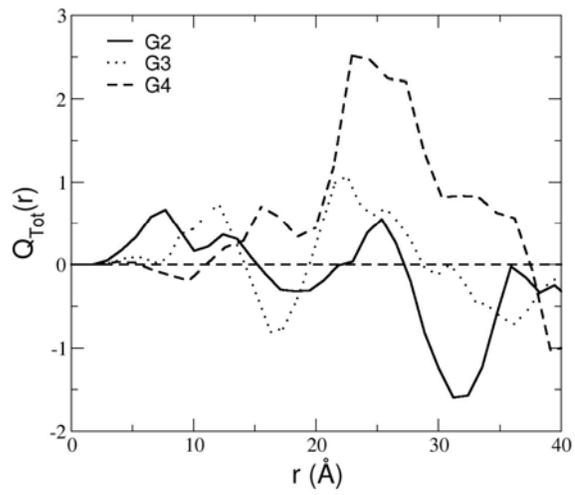

**Figure 3:**

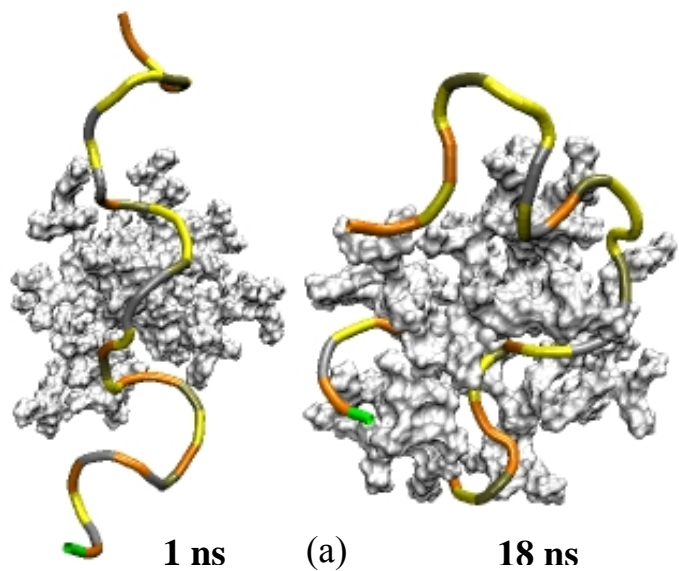

**1 ns**  (a)  **18 ns**

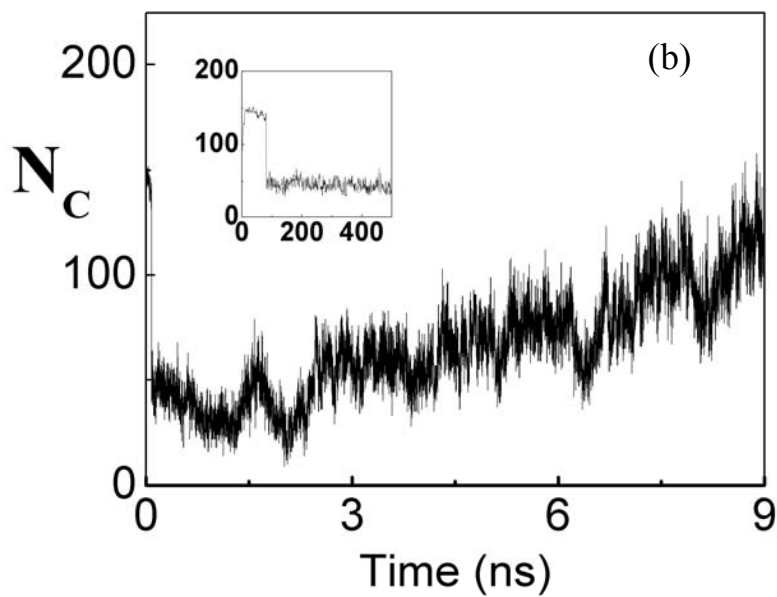

**Figure 4**

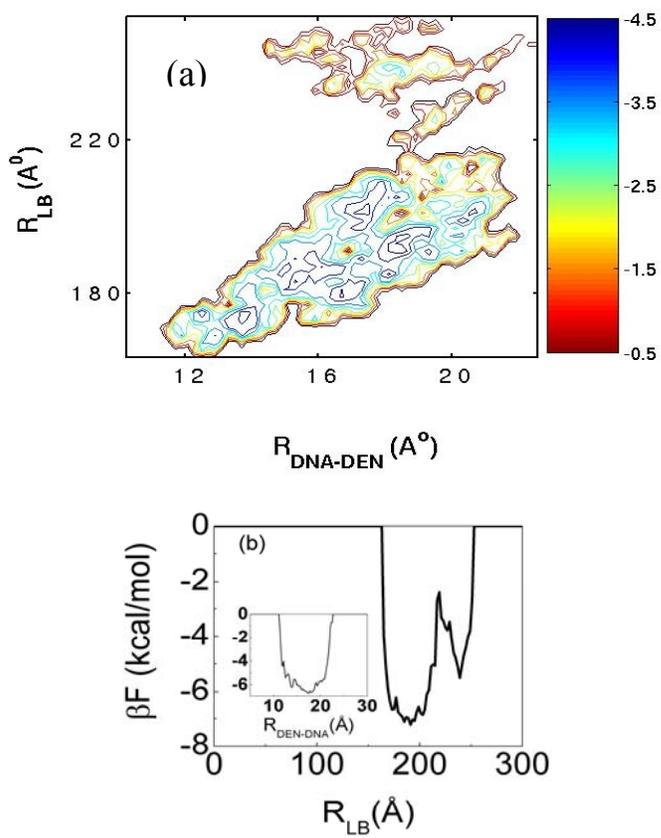

Figure 5

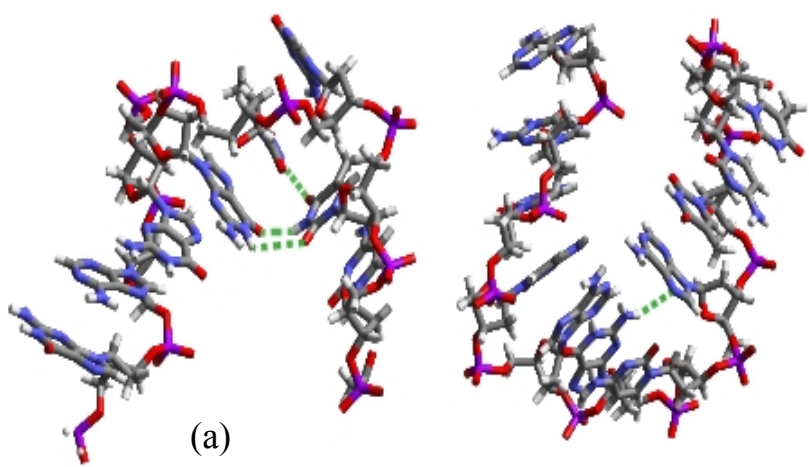

(a)

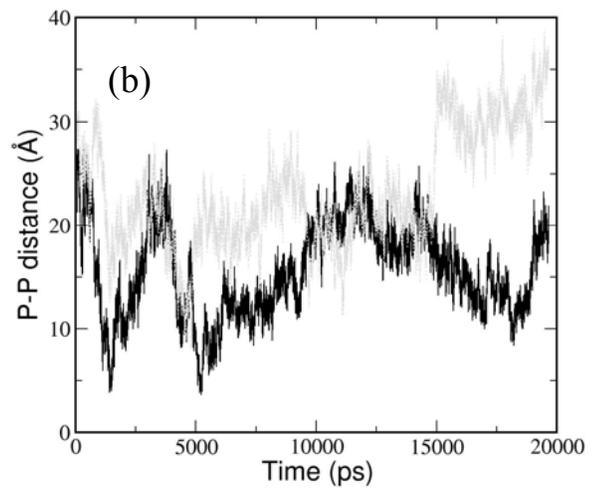

(b)

Figure 6

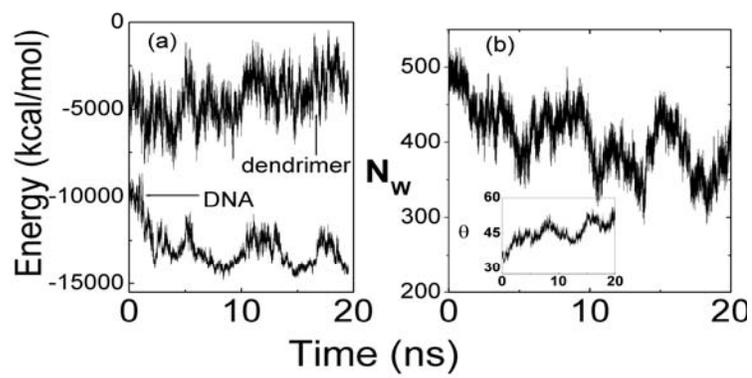

Figure 7

Table 1:

**Entropy of the homopolymer in the free as well as in bound states. Binding Energy for dna and dendrimer for various homopolymer sequence.**

| Sequence | Entropy (kcal/mol) | |
|---|---|---|
| | Free | Complex |
| polyA | 881.5 | 872.58 |
| polyT | 936.82 | 888.68 |
| PolyG | 909.28 | 946.7 |
| PolyC | 824.41 | 826.54 |

| Complex | ΔH (kcal/mol) | ΔG (kcal/mol) |
|---|---|---|
| G4+polyA | -85.2 | 79.18 |
| G4+polyT | -2.3 | 634.05 |
| G4+PolyG | -135.77 | -408.18 |
| G4+PolyC | -85.3 | -362.1 |